\documentclass[twocolumn,prl,a4paper,superscriptaddress,showpacs,preprintnumbers,tightenlines]{revtex4}
\usepackage{amsmath}
\usepackage{graphicx}
\usepackage{epsfig}

\usepackage{color}
\usepackage{dcolumn}
\newcolumntype{B}{D{B}{}{-1}}

\begin{document}

\setlength{\unitlength}{1pt}

\title{ \quad\\[1.0cm] Measurement of {\boldmath $\Upsilon$}(5S) decays 
to {\boldmath $B^0$} and {\boldmath $B^+$} mesons}

\affiliation{Budker Institute of Nuclear Physics, Novosibirsk}
\affiliation{Faculty of Mathematics and Physics, Charles University, Prague}
\affiliation{University of Cincinnati, Cincinnati, Ohio 45221}
\affiliation{Justus-Liebig-Universit\"at Gie\ss{}en, Gie\ss{}en}
\affiliation{Gyeongsang National University, Chinju}
\affiliation{Hanyang University, Seoul}
\affiliation{University of Hawaii, Honolulu, Hawaii 96822}
\affiliation{High Energy Accelerator Research Organization (KEK), Tsukuba}
\affiliation{Hiroshima Institute of Technology, Hiroshima}
\affiliation{Institute of High Energy Physics, Chinese Academy of Sciences, Beijing}
\affiliation{Institute of High Energy Physics, Vienna}
\affiliation{Institute of High Energy Physics, Protvino}
\affiliation{INFN - Sezione di Torino, Torino}
\affiliation{Institute for Theoretical and Experimental Physics, Moscow}
\affiliation{J. Stefan Institute, Ljubljana}
\affiliation{Kanagawa University, Yokohama}
\affiliation{Institut f\"ur Experimentelle Kernphysik, Karlsruhe Institut f\"ur Technologie, Karlsruhe}
\affiliation{Korea University, Seoul}
\affiliation{Kyungpook National University, Taegu}
\affiliation{\'Ecole Polytechnique F\'ed\'erale de Lausanne (EPFL), Lausanne}
\affiliation{Faculty of Mathematics and Physics, University of Ljubljana, Ljubljana}
\affiliation{University of Maribor, Maribor}
\affiliation{Max-Planck-Institut f\"ur Physik, M\"unchen}
\affiliation{University of Melbourne, School of Physics, Victoria 3010}
\affiliation{Nagoya University, Nagoya}
\affiliation{Nara Women's University, Nara}
\affiliation{National Central University, Chung-li}
\affiliation{Department of Physics, National Taiwan University, Taipei}
\affiliation{H. Niewodniczanski Institute of Nuclear Physics, Krakow}
\affiliation{Nippon Dental University, Niigata}
\affiliation{Niigata University, Niigata}
\affiliation{University of Nova Gorica, Nova Gorica}
\affiliation{Novosibirsk State University, Novosibirsk}
\affiliation{Osaka City University, Osaka}
\affiliation{Panjab University, Chandigarh}
\affiliation{University of Science and Technology of China, Hefei}
\affiliation{Seoul National University, Seoul}
\affiliation{Sungkyunkwan University, Suwon}
\affiliation{School of Physics, University of Sydney, NSW 2006}
\affiliation{Tata Institute of Fundamental Research, Mumbai}
\affiliation{Excellence Cluster Universe, Technische Universit\"at M\"unchen, Garching}
\affiliation{Tohoku Gakuin University, Tagajo}
\affiliation{Tohoku University, Sendai}
\affiliation{Department of Physics, University of Tokyo, Tokyo}
\affiliation{Tokyo Metropolitan University, Tokyo}
\affiliation{Tokyo University of Agriculture and Technology, Tokyo}
\affiliation{IPNAS, Virginia Polytechnic Institute and State University, Blacksburg, Virginia 24061}
\affiliation{Yonsei University, Seoul}
  \author{A.~Drutskoy}\affiliation{University of Cincinnati, Cincinnati, Ohio 45221} 
  \author{I.~Adachi}\affiliation{High Energy Accelerator Research Organization (KEK), Tsukuba} 
  \author{H.~Aihara}\affiliation{Department of Physics, University of Tokyo, Tokyo} 
  \author{V.~Aulchenko}\affiliation{Budker Institute of Nuclear Physics, Novosibirsk}\affiliation{Novosibirsk State University, Novosibirsk} 
  \author{T.~Aushev}\affiliation{\'Ecole Polytechnique F\'ed\'erale de Lausanne (EPFL), Lausanne}\affiliation{Institute for Theoretical and Experimental Physics, Moscow} 
  \author{A.~M.~Bakich}\affiliation{School of Physics, University of Sydney, NSW 2006} 
  \author{V.~Balagura}\affiliation{Institute for Theoretical and Experimental Physics, Moscow} 
  \author{V.~Bhardwaj}\affiliation{Panjab University, Chandigarh} 
  \author{M.~Bischofberger}\affiliation{Nara Women's University, Nara} 
  \author{A.~Bondar}\affiliation{Budker Institute of Nuclear Physics, Novosibirsk}\affiliation{Novosibirsk State University, Novosibirsk} 
% ...
\collaboration{The Belle Collaboration}



\begin{abstract}
Decays of the $\Upsilon$(5S) resonance to channels with $B^+$ and $B^0$ mesons
are studied using a 23.6\,fb$^{-1}$ data sample
collected with the Belle
detector at the KEKB asymmetric-energy $e^+ e^-$ collider.
Fully reconstructed $B^+ \to J/\psi K^+$, $B^0 \to J/\psi K^{*0}$,
$B^+ \to \bar{D}^0 \pi^+$ and $B^0 \to D^- \pi^+$ decays are used
to obtain the charged and neutral $B$ production rates per $b\bar{b}$ event,
$f(B^+) = (72.1 ^{+3.9}_{-3.8} \pm 5.0)\%$ and
$f(B^0) = (77.0 ^{+5.8}_{-5.6} \pm 6.1)\%$.
Assuming equal rates to $B^+$ and $B^0$ mesons
in all channels produced at the $\Upsilon$(5S) energy,
we measure the fractions for transitions to
two-body and three-body channels with $B$ meson pairs,
$f(B\bar{B}) = (5.5\,^{+1.0}_{-0.9} \pm 0.4)\,\%$,
$f(B\bar{B}^*+B^*\bar{B}) = (13.7 \pm 1.3 \pm 1.1)\,\%$,
$f(B^*\bar{B}^*) = (37.5\,^{+2.1}_{-1.9} \pm 3.0)\,\%$,
$f(B\bar{B}\,\pi) = (0.0 \pm 1.2 \pm 0.3)\,\%$,
$f(B\bar{B}^\ast\pi+B^\ast\bar{B}\pi) = (7.3\,^{+2.3}_{-2.1} \pm 0.8)\,\%$,
and $f(B^\ast\bar{B}^\ast\pi) = (1.0\,^{+1.4}_{-1.3} \pm 0.4)\,\%$.
The latter three fractions are obtained assuming isospin conservation.
\end{abstract}

\pacs{13.25.Gv, 13.25.Hw, 14.40.Pq, 14.40.Nd}

\maketitle


{\renewcommand{\thefootnote}{\fnsymbol{footnote}}}
\setcounter{footnote}{0}

New aspects of beauty dynamics can be explored
using the large data sample recently collected by Belle
at the center-of-mass (CM) energy of the $\Upsilon$(5S) 
resonance (also referred to as $\Upsilon$(10860)).
Decays of bottomonium states with masses higher
than the mass of the $\Upsilon$(4S) are almost unexplored with only
a very limited number of experimental and theoretical studies to date.
At the $\Upsilon$(5S) energy a $b\bar{b}$ quark pair can be produced and 
hadronize into various final states,
which can be classified as
two-body $B_s^0$ ($B_s^0\bar{B}_s^0$, 
$B_s^0\bar{B}_s^\ast$, $B_s^\ast\bar{B}_s^0$, $B_s^\ast\bar{B}_s^\ast$)
\cite{cleoi,beli},
two-body $B$ 
($B\bar{B}$, $B\bar{B}^\ast$, $B^\ast\bar{B}$, $B^\ast\bar{B}^\ast$),
three-body ($B\bar{B}\,\pi$, $B\bar{B}^\ast\,\pi$, 
$B^\ast\bar{B}\,\pi$, $B^\ast\bar{B}^\ast\,\pi$), and 
four-body ($B\bar{B}\,\pi \pi$) channels.
Here $B$ denotes a $B^+$ or $B^0$ meson and 
$\bar{B}$ denotes a $B^-$ or $\bar{B}^0$ meson.
The excited states decay to their ground states via
$B^\ast \to B\gamma$ and $B_s^\ast \to B_s^0\gamma$.
Moreover, a $b\bar{b}$ quark pair can also hadronize to a bottomonium
state accompanied by $\pi$, $K$ or $\eta$ mesons, for example through
a $\Upsilon$(5S)$\to \Upsilon$(1S)$\,\pi^+\pi^-$ decay \cite{chen}.
In addition, initial-state radiation (ISR) can affect the final
states listed above and must be taken into account \cite{isr}.
Fractions for all of these channels
provide important information about $b$-quark dynamics.

The first study of $B$ production at the
$\Upsilon$(5S) was performed by CLEO \cite{cleoc,cleob}
using a $0.42$\,fb$^{-1}$ data sample. They found
the fraction of events with $B^{+/0}$ pairs
to be $(58.9 \pm 10.0 \pm 9.2)\%$.
CLEO interpreted the
remaining $(41.1 \pm 10.0 \pm 9.2)\%$ as the fraction of events
with $B_s^0$ mesons. Within the large uncertainties
this fraction is 1.6$\sigma$ larger than the $B_s^0$ event 
fraction $f_s = (18.0 \pm 1.3 \pm 3.2)\%$,
directly measured by Belle \cite{beli}.
Both values for this fraction were obtained assuming that
contributions from channels with bottomonium states are negligibly small. 
Among the $B^{+/0}$ pair events
CLEO found the two-body fractions for the $B^\ast\bar{B}^\ast$
and $B\bar{B}^\ast + B^\ast\bar{B}$ channels to be
$(74 \pm 15 \pm 8)\%$ and $(24 \pm 9 \pm 3)\%$ \cite{cleoc}, respectively.
The $B\bar{B}$ channel and multibody channels were not observed and
corresponding upper limits were set.

Several theoretical papers have been devoted to $\Upsilon$(5S) decays 
to final states
with two-body $B_s^0$ and $B^{+/0}$ pairs \cite{teoa,teob,teoc,teod}.
The $B^\ast\bar{B}^\ast$ channel is predicted to be dominant
with the fraction over all $b\bar{b}$ events 
within the range $(30-69)\%$ \cite{teoc,teod}.
In these model calculations the other two possible channels have
smaller fractions with predictions covering a broad range.
Multibody channels have also been theoretically studied \cite{sim,lel};
the three-body fractions are found to be about two or three 
orders of magnitude smaller than the two-body fractions.
Interesting information about a possible gluonic component of 
the $\Upsilon$(5S) can also be obtained 
from measurements of the three-body decays \cite{est},
if 200 or more events can be reconstructed in 
a three-body channel.

Here we fully reconstruct the decay modes
$B^+ \to J/\psi K^+$, $B^0 \to J/\psi K^{*0}$,
$B^+ \to \bar{D}^0 \pi^+$ (using two $\bar{D}^0$ modes), 
and $B^0 \to D^- \pi^+$.
Charge-conjugate modes are implicitly included throughout this work.
The reconstructed modes have large and precisely measured branching 
fractions \cite{pdg} and contain only charged particles.

The data were collected with the Belle detector \cite{belle}
at KEKB \cite{kekb}, an asymmetric-energy double storage ring $e^+\,e^-$
collider.
This analysis is based on a sample of $23.6\,\mathrm{fb}^{-1}$
taken at the $\Upsilon$(5S) CM energy of $\sim$10867 MeV
and containing 
$N_{b\bar{b}}^{\Upsilon{\rm (5S)}} = (7.13 \pm 0.34) \times 10^6$
produced events \cite{beli}.
The Belle detector is a general-purpose magnetic spectrometer 
described in detail elsewhere~\cite{belle}.

Charged tracks are assigned as pions or kaons based on
a likelihood ratio 
${\cal L}_{K/\pi} = {\cal L}_K/({\cal L}_K + {\cal L}_{\pi})$, 
which includes information obtained from the three Belle  
particle identification detector subsystems~\cite{belle}.
The identification efficiency for particles
used in this analysis varies from 85$\%$ to 92$\%$ (91$\%$ to 98$\%$)
for kaons (pions). 
The electron and muon identification requirements 
are described in Refs. \cite{ele,muo}.

The $K^{*0}$, $\bar{D}^0$, $D^-$ and $J/\psi$ candidates are reconstructed
in the $K^{*0} \to K^+ \pi^-$, $\bar{D}^0 \to K^+ \pi^-$, 
$\bar{D}^0 \to K^+ \pi^+ \pi^- \pi^-$, $D^- \to K^+ \pi^- \pi^-$,
$J/\psi \to e^+ e^-$ and $J/\psi \to \mu^+ \mu^-$ modes.
We require the invariant masses to be in the following intervals
around the nominal masses: $\pm 50\,$MeV/$c^2$ for $K^{*0}$,
$\pm 10\,$MeV/$c^2$ for $\bar{D}^0$ and $D^-$,
$\pm 30\,$MeV/$c^2$ for $J/\psi \to \mu^+ \mu^-$, 
and $^{+30}_{-100}\,$MeV/$c^2$ for $J/\psi \to e^+ e^-$.
A vertex- and mass-constrained fit is applied to 
$J/\psi$, $\bar{D}^0$ and $D^-$ candidates to improve the $B$ signal
resolution.

$B$ decays are fully reconstructed and identified using two variables:
the energy difference $\Delta E\,=\,E^{\rm CM}_{B}-E^{\rm CM}_{\rm beam}$
and the beam-energy-constrained mass
$M_{\rm bc} = \sqrt{(E^{\rm CM}_{\rm beam})^2\,-\,(p^{\rm CM}_{B})^2}$,
where $E^{\rm CM}_{B}$ and $p^{\rm CM}_{B}$ are the energy and momentum
of the $B$ candidate in the $e^+ e^-$ CM system,
and $E^{\rm CM}_{\rm beam}$ is the CM beam energy.
The intermediate two-body and multibody channels with $B^{+/0}$ pairs
cluster in distinct regions of the $M_{\rm bc}$ and $\Delta E$ plane. 
However, all channels are
distributed along a straight line described approximately 
by the function \mbox{$\Delta E = m_B-M_{\rm bc}$}, where $m_B$ is
the nominal $B$ mass (fixed to $5.28\,$GeV/$c^2$).

After all selections the dominant background is from
$e^+e^- \rightarrow q \bar{q}$ continuum events ($q = u,d,s,$ or $c$).
Events with $B$ mesons tend to be spherical,
whereas continuum events are expected to be jet-like.
To suppress continuum background, we apply topological cuts.
The ratio of the second to the zeroth Fox-Wolfram moments \cite{fox}
is required to be less than 0.5 for the low background final states
with a $J/\psi$, and less than 0.4 for all others.
The angle in the CM
between the thrust axis of the particles forming the $B$ candidate
and the thrust axis of all other particles in the event, 
$\theta^*_{\rm thr}$, must satisfy $|{\rm cos}\,\theta^*_{\rm thr}| < 0.9$
for the final states with a $J/\psi$, and 
$|{\rm cos}\,\theta^*_{\rm thr}| < 0.75$ for all others.
More than one $B^{+/0}$ candidate per event is allowed,
however the probability of multiple candidates is less than 1$\%$ 
for the modes used here, where the final states
contain charged particles only.

The two-dimensional $M_{\rm bc}$ and $\Delta E$ scatter plots
for the $B^+ \to J/\psi K^+$, $B^0 \to J/\psi K^{*0}$,
$B^+ \to \bar{D}^0 \pi^+$ ($\bar{D}^0 \to K^+ \pi^-$ and 
$\bar{D}^0 \to K^+ \pi^+ \pi^- \pi^-$)
and $B^0 \to D^- \pi^+$ modes are obtained and
are shown in Fig.~1. Events are clearly
concentrated along the line $\Delta E = m_B - M_{\rm bc}$
corresponding to $B$ pair production.
The signal regions shown in Fig.~1 are restricted to a $\pm 30\,$MeV 
interval in $\Delta E$ (corresponding to (2.5--4.0)$\,\sigma$
for the studied modes) and to the kinematically allowed
$5.268\,$GeV/$c^2 < M_{\rm bc} < 5.440\,$GeV/$c^2$ range.
The location of specific channels within the signal bands
will be discussed below and is shown in Fig.~2a.

\begin{figure}[h!]
\vspace{-0.1cm}
\begin{center}
\epsfig{file=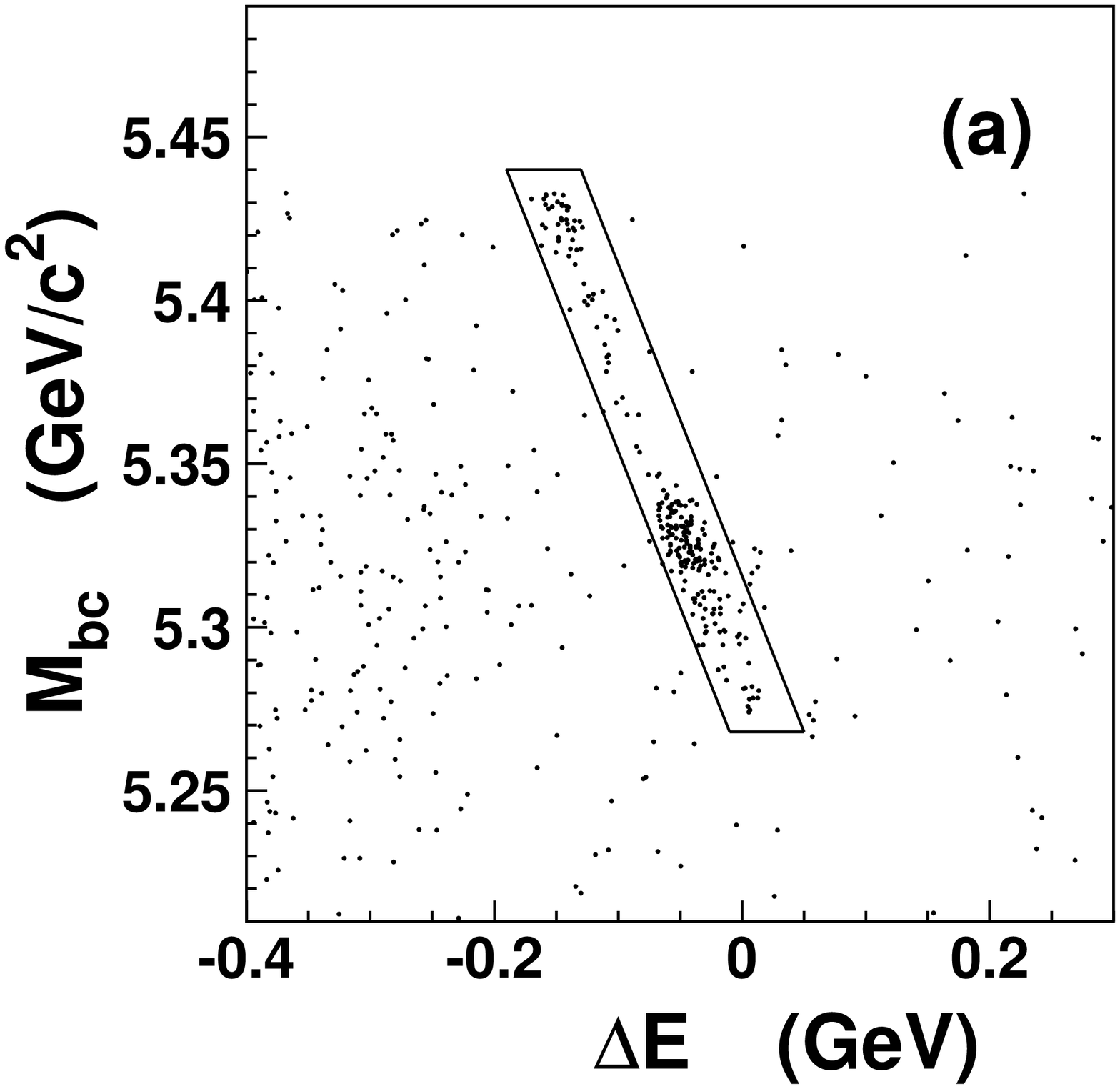,width=4.4cm,height=4.4cm}\epsfig{file=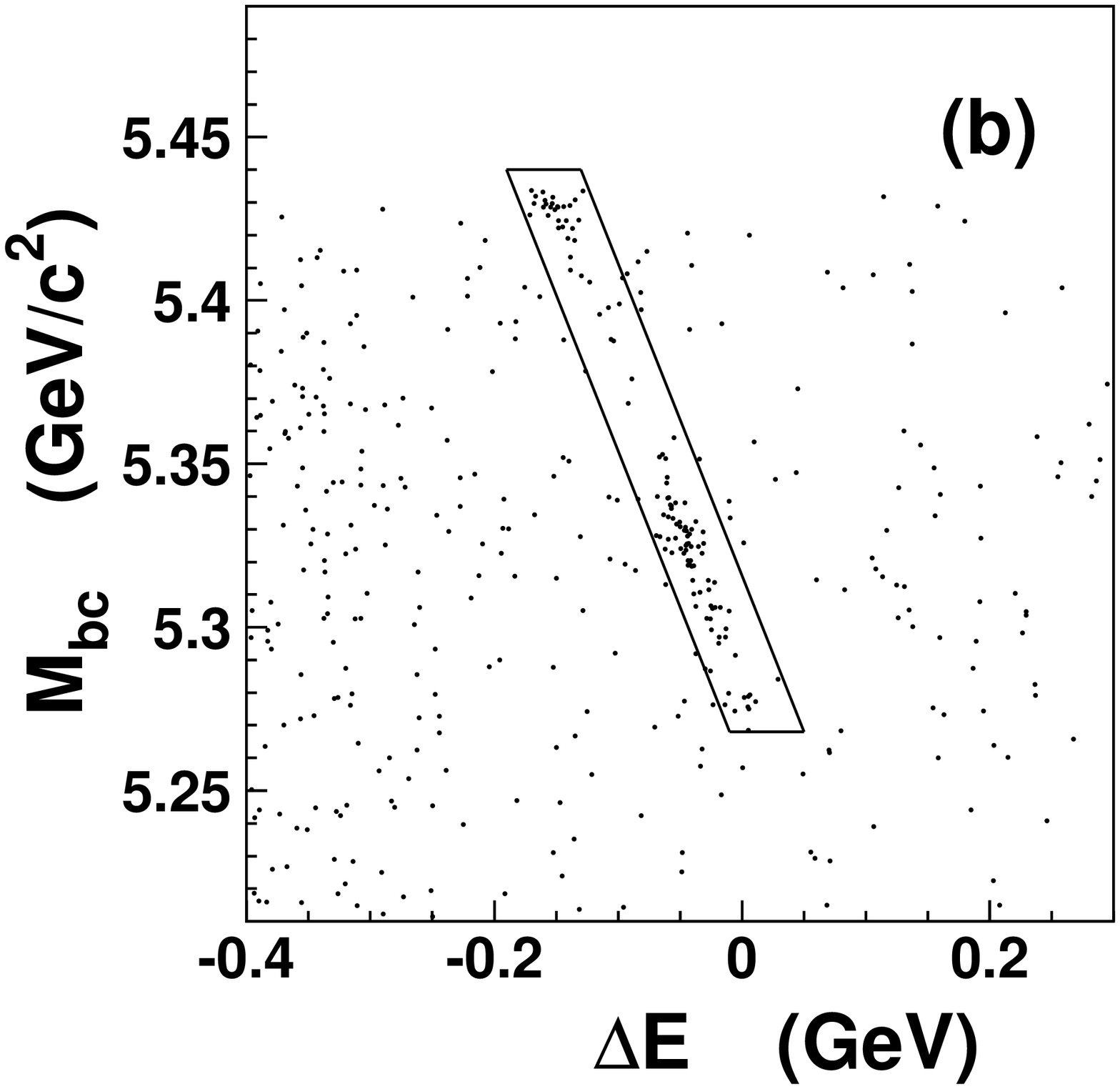,width=4.4cm,height=4.4cm}
\epsfig{file=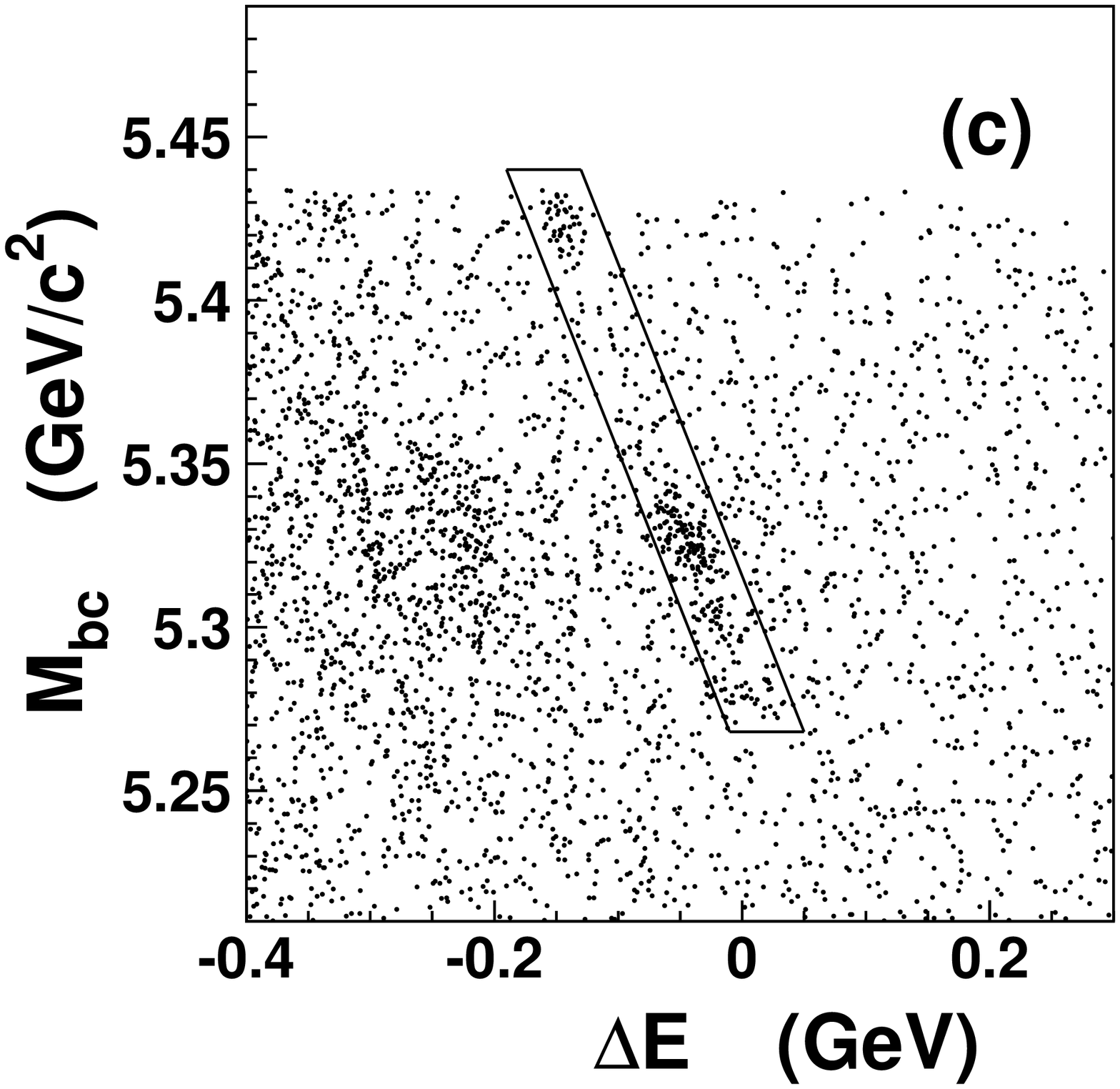,width=4.4cm,height=4.4cm}\epsfig{file=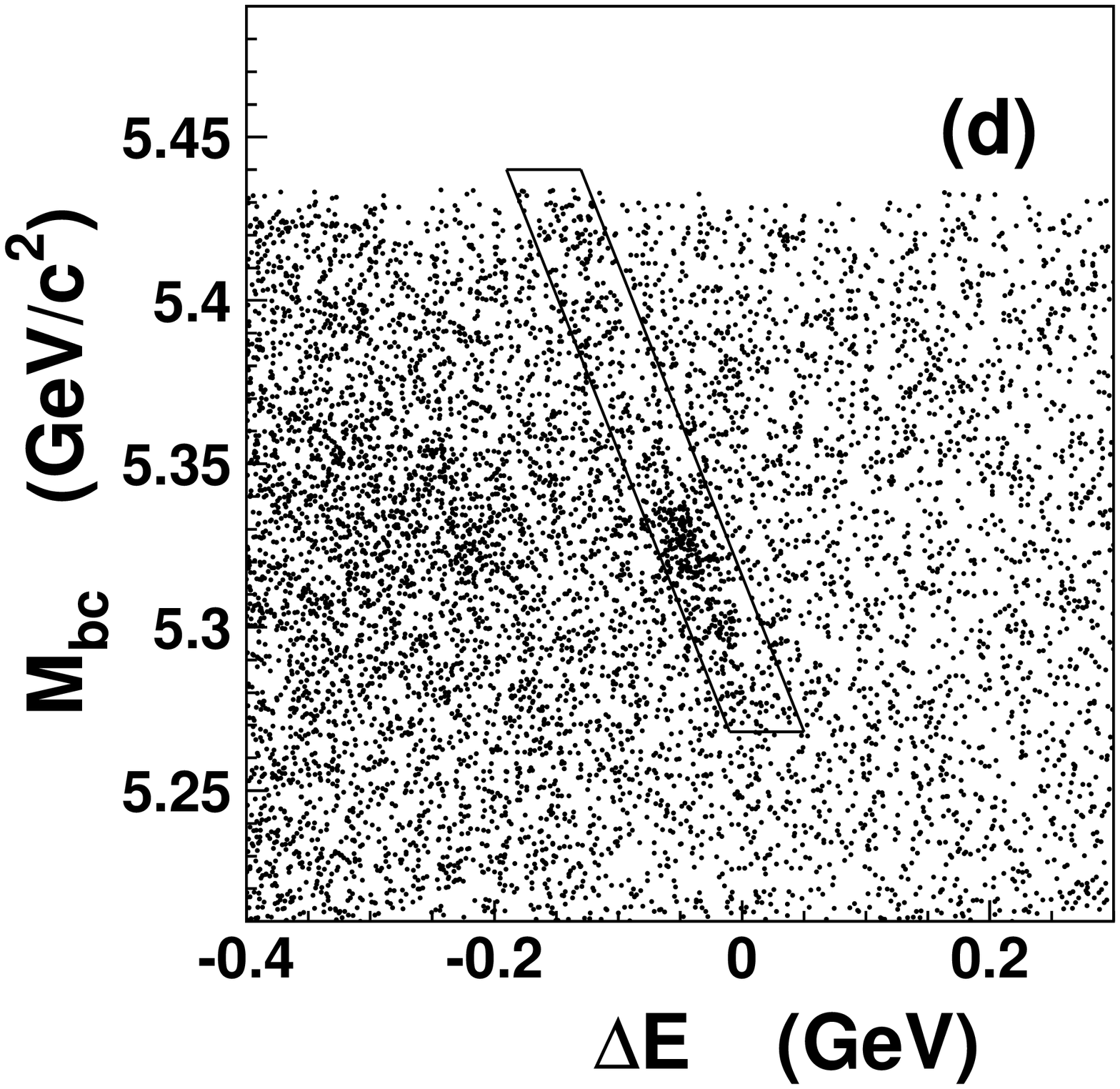,width=4.4cm,height=4.4cm}
\epsfig{file=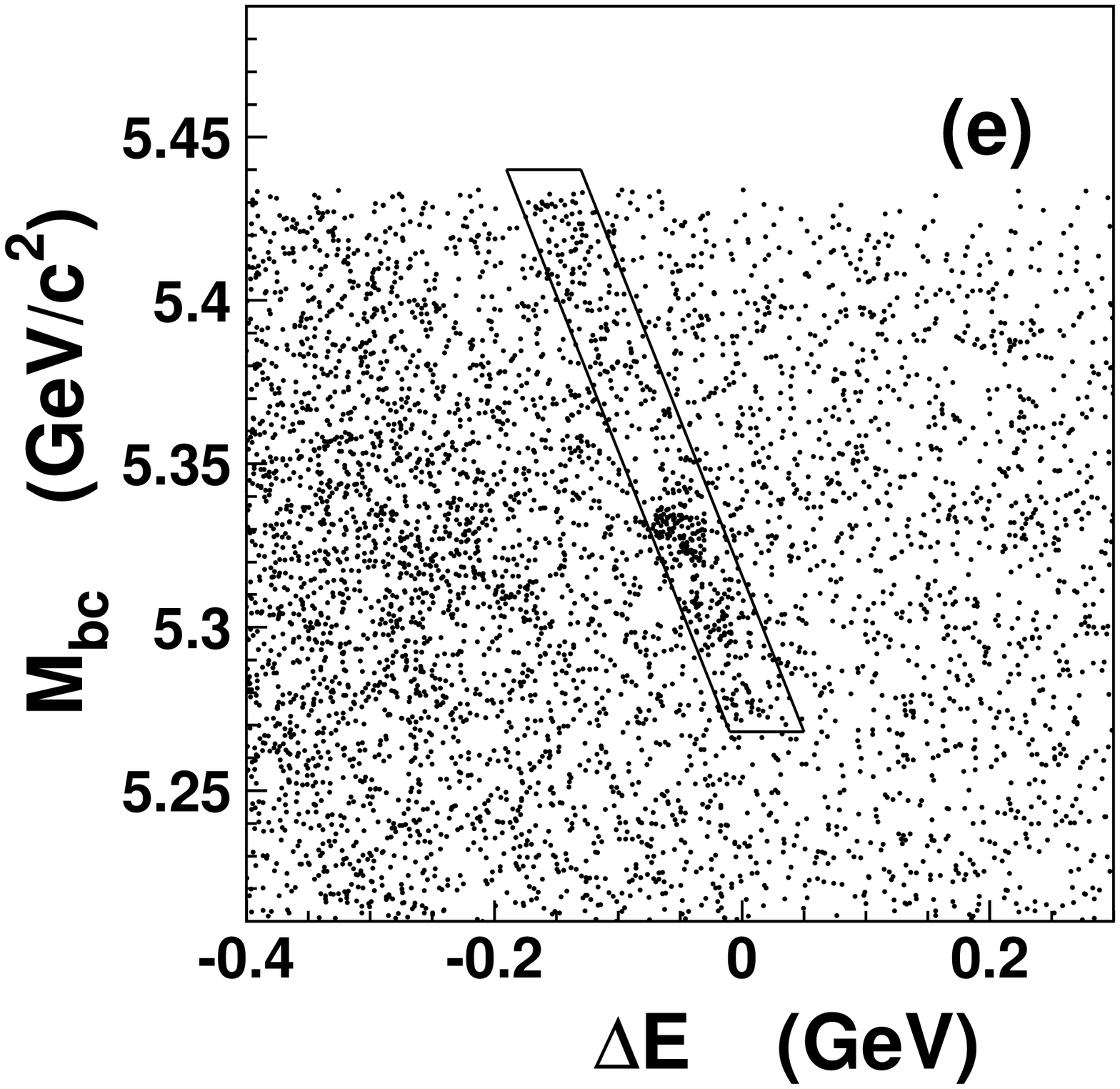,width=4.4cm,height=4.4cm}
\end{center}
\vspace{-0.6cm}
\caption{The $M_{\rm bc}$ and $\Delta E$ scatter plots
for (a) $B^+ \to J/\psi K^+$, (b) $B^0 \to J/\psi K^{*0}$,
(c) $B^+ \to \bar{D}^0 (K^+ \pi^-) \pi^+$,
(d) $B^+ \to \bar{D}^0 (K^+ \pi^+ \pi^- \pi^-) \pi^+$,
and (e) $B^0 \to D^- \pi^+$ decays. 
The bands indicate the signal regions corresponding
to the intervals $5.268\,$GeV/$c^2 < M_{\rm bc} < 5.440\,$GeV/$c^2$ and 
$|\Delta E+M_{\rm bc}-m_B| < 0.03\,$GeV.}
\vspace{-0.1cm}
\label{fig1}
\end{figure}

We use inclined \mbox{$\Delta E+M_{\rm bc}-m_B$} projections of the
two-dimensional scatter plots for all events within the range 
$5.268\,$GeV/$c^2 < M_{\rm bc} < 5.440\,$GeV/$c^2$
to obtain integrated $B$ decay event yields.
We fit these distributions
with a function including two terms:
a Gaussian to describe the signal and a first-order polynomial
to describe background. The regions where cross-channel $B$
backgrounds can contribute are not used in the fits.

Using the fit results, the charged and neutral $B$ production rates 
per $b\bar{b}$ event are obtained from the formula:
\begin{equation}
f(B^{+/0}) \ = \ Y^{\rm fit}_{B \to X} \ / \ ( N_{b\bar{b}}^{\Upsilon{\rm (5S)}} \times \epsilon_{B \to X} \times {\cal B}_{B \to X}) \ , 
\vspace{-0.1cm}
\end{equation} 
where $Y^{\rm fit}_{B \to X}$ is the event yield
obtained from the fit for a specific mode $B \to X$,
$\epsilon_{B \to X}$ is the  reconstruction efficiency 
including intermediate branching fractions, 
and ${\cal B}_{B \to X}$ is the corresponding $B$ decay branching
fraction \cite{pdg}.
The event yields, efficiencies, and  production rates are listed in Table I.

\renewcommand{\arraystretch}{1.3}
\begin{table}[h!]
\vspace{-0.3cm}
\caption{Event yields obtained from fits, efficiencies, 
and $B$ production rates $f(B^{+/0})$. Efficiencies include
intermediate $J/\psi$, $K^{*0}$, $\bar{D}^0$ and $D^-$ branching fractions.}
\begin{center}
\vspace{-0.1cm}
\begin{tabular}
{@{\hspace{0.03cm}}l@{\hspace{0.1cm}} @{\hspace{0.1cm}}c@{\hspace{0.1cm}} @{\hspace{0.05cm}}c@{\hspace{0.05cm}} @{\hspace{0.1cm}}c@{\hspace{0.03cm}}}
\hline \hline
 Decay mode & Yield & Efficiency,$\,\%$ & $f(B^{+/0})$,$\,\%$ \\ 
\hline
$B^+ \to J/\psi K^+$ &  $221^{+16}_{-15}$ & $3.41$ & $89.0^{+6.3}_{-6.1} \pm 8.0$  \\
$B^0 \to J/\psi K^{*0}$ &  $105 \pm 11$ & $1.30$ & $85.3^{+9.2}_{-8.8} \pm 8.8$ \\
$B^+ \to \bar{D}^0 (K \pi) \pi^+$ &  $215 \pm 21$ & $0.97$ & $64.0 \pm 6.2 \pm 4.9$ \\
$B^+ \to \bar{D}^0 (K 3 \pi) \pi^+$ &  $275 \pm 32$ & $1.17$ & $68.3^{+8.0}_{-8.1} \pm 6.4$ \\
$B^0 \to D^- \pi^+$ &  $247 \pm 25$ & $1.80$ & $72.9 \pm 7.4 \pm 6.4$ \\
\hline \hline
\end{tabular}
\end{center}
\vspace{-0.6cm}
\end{table}

The systematic uncertainties include those due to
the determination of the number of $b\bar{b}$ events (4.7$\%$),
charged track reconstruction efficiency (1$\%$ per track),
particle identification (0.5--1.0$\%$ per $\pi$ and $K$,
2$\%$ per electron, 3$\%$ per muon),
$J/\psi$, $\bar{D}^0$ and $D^-$ mass cut efficiencies (2$\%$),
signal and background modeling in the fit procedure (2$\%$),
MC statistics in efficiency determination (1--2$\%$),
shape of the $B$ meson angular distribution relative to the beam axis
direction in the CM system (1$\%$),
and PDG branching fractions (3--5$\%$). 
All systematic uncertainties are combined in quadrature
to obtain the total systematic uncertainty.
The same set of uncertainties is used below to obtain the 
total systematic uncertainties for values averaged over several
$B$ modes, however the correlated and uncorrelated uncertainties 
are treated separately in this case. 
All uncorrelated uncertainties
from all $B$ modes are varied individually to obtain their contribution
to the total systematic uncertainty of the averaged value, while 
correlated uncertainties are varied simultaneously for all channels.

Using the rates shown in Table 1, the average production rates
$f(B^+) = (72.1 ^{+3.9}_{-3.8} \pm 5.0)\%$ and 
$f(B^0) = (77.0^{+5.8}_{-5.6} \pm 6.1)\%$ are obtained. 
As expected, these rates are equal within uncertainties.
The average of charged and neutral $B$ modes is 
$(73.7 \pm 3.2 \pm 5.1)\%$.
Within uncertainties this rate is consistent
with the CLEO value of $(58.9 \pm 10.0 \pm 9.2)\%$ \cite{cleob}.

Since the $f(B^+)$ and $f(B^0)$ values are consistent with
isospin symmetry, it is reasonable to assume
that equal numbers of $B^+$ and $B^0$ mesons are produced in all 
possible channels.
Therefore the five $B$ decay modes are treated simultaneously
everywhere below.
First, events in the signal bands shown in Fig.~1 are projected 
onto $M_{\rm bc}$. 
To describe combinatorial backgrounds under the signal, 
we define sidebands, which have  the same shape
as the signal region, but are shifted by $\pm$70~MeV in $\Delta E$,
and project similarly onto $M_{\rm bc}$.
Assuming that background is distributed linearly in $\Delta E$,
we model the combinatorial background by taking the average of the two
sideband $M_{\rm bc}$ distributions. 

Figure 2(a) shows the $M_{\rm bc}$ distributions for the possible two-, three- 
and four-body channels obtained from MC simulation for the 
$B^0 \to D^- \pi^+$ decay. 
The three-body and four-body channels
are simulated assuming a pure phase-space decay model.
Figure 2(a) shows that the two-body channels are well separated. 
The three-body $B\bar{B}\,\pi$, $B\bar{B}^\ast\,\pi+B^\ast\bar{B}\,\pi$
and $B^\ast\bar{B}^\ast\,\pi$ channels 
have broader distributions
in the higher mass region $M_{\rm bc}>\,$5.35~GeV/$c^2$.
The four-body $B\bar{B}\,\pi \pi$ channel is the cross-hatched peak 
on the right side of Fig.~2(a). 

\begin{figure}[h!]
\vspace{-0.3cm}
\begin{center}
\epsfig{file=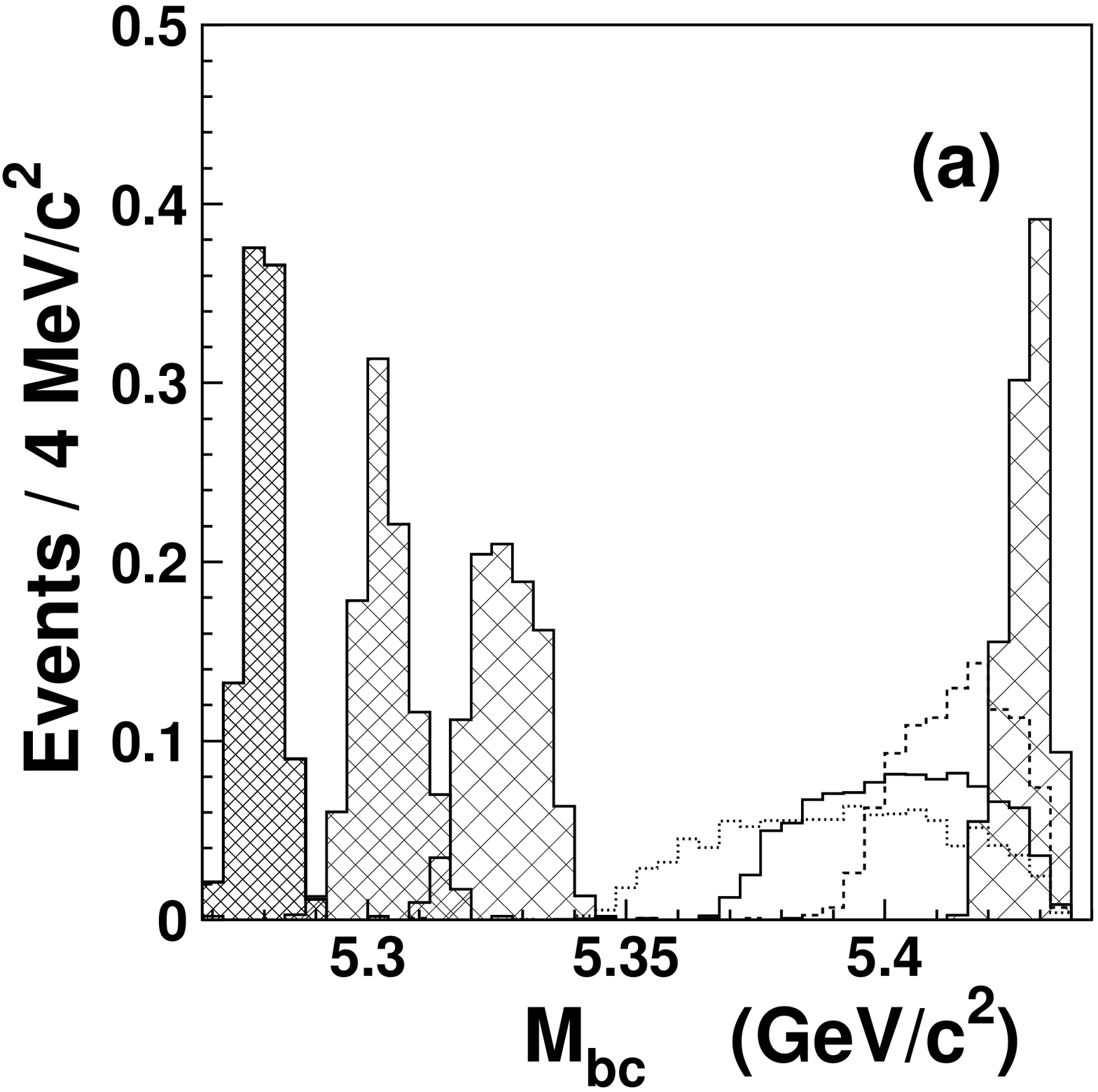,width=4.4cm,height=4.4cm}\epsfig{file=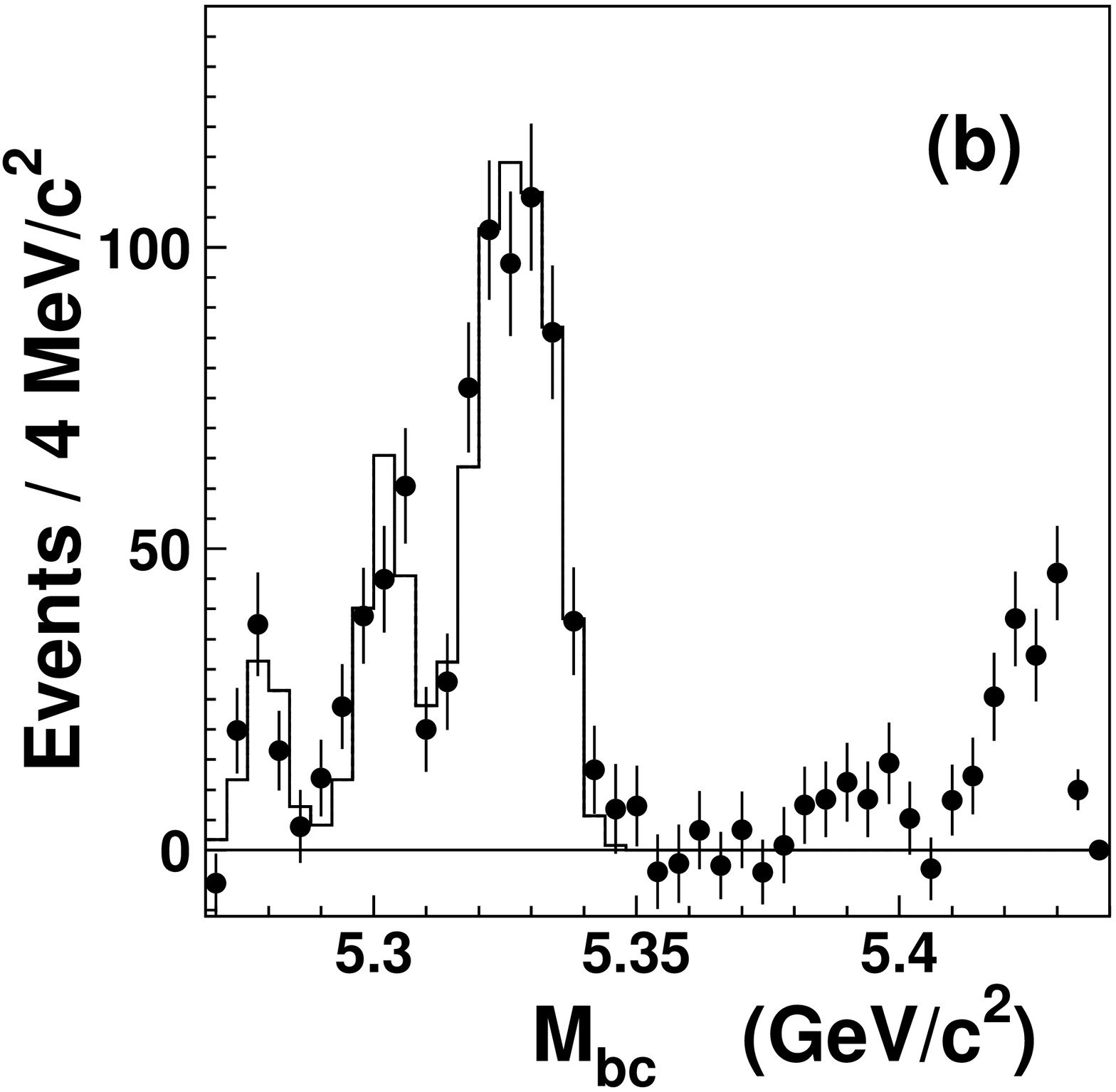,width=4.4cm,height=4.4cm}
\end{center}
\vspace{-0.6cm}
\caption{(a) MC simulated $M_{\rm bc}$ distributions for the 
$B^0 \to D^- \pi^+$ decay for 
$B\bar{B}$, $B\bar{B}^\ast+B^\ast\bar{B}$, $B^\ast\bar{B}^\ast$ and
$B\bar{B}\,\pi \pi$ channels
(cross-hatched histograms from left to right), and also for the three-body
channels $B\bar{B}^\ast\,\pi+B^\ast\bar{B}\,\pi$ (plain histogram),
$B\bar{B}\,\pi$ (dotted) and $B^\ast\bar{B}^\ast\,\pi$ (dashed).
The distributions are normalized to unity. 
(b) $M_{\rm bc}$ distribution in data after background subtraction.
The sum of the five studied $B$ decays (points with error bars) 
and results of the
fit (histogram) used to extract the two-body channel fractions are shown.
The fitting procedure is described in the text.}  
\vspace{-0.05cm}
\end{figure}

The matrix elements responsible
for the three- and four-body decays are not known, and
the rates of the three- and four-body contributions
cannot be obtained in a model-independent way from a fit to these 
$M_{\rm bc}$ distributions.
We therefore restrict the fit 
to the region $5.268\,$GeV/$c^2\,<\,M_{\rm bc}\,<$\,5.348~GeV/$c^2$
to extract the two-body channel fractions.
To obtain the sum of the other contributions in the large $M_{\rm bc}$ 
region, only the events from the interval
$5.348\,$GeV/$c^2\, <\,M_{\rm bc}\,<$\,5.440~GeV/$c^2$  are
selected. For these distributions we apply a fit procedure similar to that
used above to obtain the full $f(B^{+/0})$ production rate 
for the entire interval
$5.268\,$GeV/$c^2\, <\,M_{\rm bc}\,<$\,5.440~GeV/$c^2$.
The total signal yield, summed over all
modes and channels in this region, is $228.7\,^{+22.9}_{-22.3}$.

The fitting procedure used to extract the production rates of the
two-body decay channels treats the five signal and five sideband 
$M_{\rm bc}$ distributions simultaneously. The following four components
with fixed shapes and floating normalizations are included
in the fit for each distribution: $B\bar{B}$,
$B\bar{B}^\ast+B^\ast\bar{B}$,
$B^\ast\bar{B}^\ast$, and combinatorial background.
The shapes of the signal components are taken from MC simulation, and
those of combinatorial background are modeled by the sideband data.
The normalization parameters (channel fractions) for the two-body components
are constrained to be equal for all five studied $B$ decays. 
The sum of background-subtracted $M_{\rm bc}$ distributions
for the five studied $B$ decays is shown in Fig.~2(b), where the result of 
the likelihood fit in the region 
$5.268\,$GeV/$c^2\, <\,M_{\rm bc}\,<$5.348~GeV/$c^2$ 
used to obtain two-body channel fractions is superimposed.
The fractions obtained from the fit are listed in \mbox{Table II}. 
The systematic uncertainties
include all those described above as well as
uncertainties due to the $M_{\rm bc}$ signal shape modeling (3$\%$).

\renewcommand{\arraystretch}{1.3}
\vspace{-0.2cm}
\begin{table}[h!]
\caption{The fractions of two-body channels 
with $B^+$ and $B^0$ mesons. The fraction obtained from a fit to the
large $M_{\rm bc}$ region is also given.} 
\begin{center}
\vspace{0.2cm}
\begin{tabular}
{@{\hspace{0.3cm}}l@{\hspace{0.3cm}} @{\hspace{0.3cm}}c@{\hspace{0.3cm}}}
\hline \hline
 Channel & Fraction, $\%$ \\
\hline
$B\bar{B}$ & $5.5\,^{+1.0}_{-0.9} \pm 0.4$  \\
$B\bar{B}^\ast$+$B^\ast\bar{B}$ & $13.7 \pm 1.3 \pm 1.1$  \\
$B^\ast\bar{B}^\ast$ & $37.5\,^{+2.1}_{-1.9} \pm 3.0$  \\
Large $M_{\rm bc}$ & $17.5\,^{+1.8}_{-1.6} \pm 1.3$  \\
\hline \hline
\end{tabular}
\end{center}
\vspace{-0.5cm}
\end{table}

To reconstruct the three-body channels, we 
look for an additional charged pion produced directly in 
the $B^{(*)}\bar{B}^{(*)}\,\pi^+$ channels. 
For each charged pion not included in the reconstructed $B$ candidate, we
form right-sign 
$B^+\,\pi^-$, $\bar{B}^0\,\pi^-$, $B^0\,\pi^+$ or $B^-\,\pi^+$
combinations. We then compute the variables 
$M_{\rm bc}^{\rm mis}$ and $\Delta E^{\rm mis}$ for the missing $B$ meson,
using the energy and momentum of the reconstructed $B\pi$ combination
in the CM:
$E({B^{\rm mis}}) = 2 E_{\rm beam}^{\rm CM} - E({B\pi})^{\rm CM}$
and $p({B^{\rm mis}}) = p({B\pi})^{\rm CM}$.
In the 
$B\bar{B}^\ast\,\pi^+ +B^\ast\bar{B}\,\pi^+$ and $B^\ast\bar{B}^\ast\,\pi^+$
channels, the $\Delta E^{\rm mis}$ value will be shifted due to
unreconstructed photons from $B^*$ decays.

Figure 3(a) shows the corrected
\mbox{$\Delta E^{\rm mis}+M_{\rm bc}^{\rm mis}-m_B$}
projections for MC simulated  
$B\bar{B}\,\pi^+$, $B\bar{B}^\ast\,\pi^+ +B^\ast\bar{B}\,\pi^+$,
$B^\ast\bar{B}^\ast\,\pi^+$, and $B\bar{B}\,\pi \pi$ events where
the $B^+ \to J/\psi K^+$ mode is generated.
The reconstructed $B$ candidates are selected from the
signal region within the intervals 
$5.37\,$GeV/$c^2\, < M_{\rm bc} < 5.44\,$GeV/$c^2$ and
$|\Delta E + M_{\rm bc} - m_B| < 0.03\,$GeV.
The value \mbox{$\Delta E^{\rm mis}+M_{\rm bc}^{\rm mis}-m_B$}
is corrected by adding to it the value $\Delta E\,+\,M_{\rm bc}\,-\,m_B$.
This does not introduce any bias for the  
original $B$ mesons, but improves the resolution, because these two  
values have partially anticorrelated uncertainties.
Figure 3a shows that the $B\bar{B}\,\pi^+$,
$B\bar{B}^\ast\,\pi^+ +B^\ast\bar{B}\,\pi^+$ and $B^\ast\bar{B}^\ast\,\pi^+$
channel contributions are well separated. 
The reconstruction efficiency for the four-body channel
(the small peak in the rightmost part of Fig.~3(a)) is small
and model dependent. Therefore, we do not include 
the four-body channel in the fit procedure described below.
The background due to random charged tracks from the unobserved
$B$ meson is also shown.

Finally, the \mbox{$\Delta E^{\rm mis}\,+\,M_{\rm bc}^{\rm mis}\,-\,m_B$}
distribution is obtained in data for the sum of the five reconstructed 
$B$ modes (Fig.~3(b)).
We fit this
distribution with a function including four terms:
three Gaussians with fixed shapes and free normalizations to describe 
the $B\bar{B}\,\pi^+$, $B\bar{B}^\ast\,\pi^+ +B^\ast\bar{B}\,\pi^+$,
and $B^\ast\bar{B}^\ast\,\pi^+$ contributions, and
a second-order polynomial to describe the background.
The central positions and widths of the Gaussians 
are obtained from fits to the MC simulated distributions
shown in Fig.~3(a) and are fixed in the fit to data.

\begin{figure}[h!]
\vspace{-0.3cm}
\begin{center}
\epsfig{file=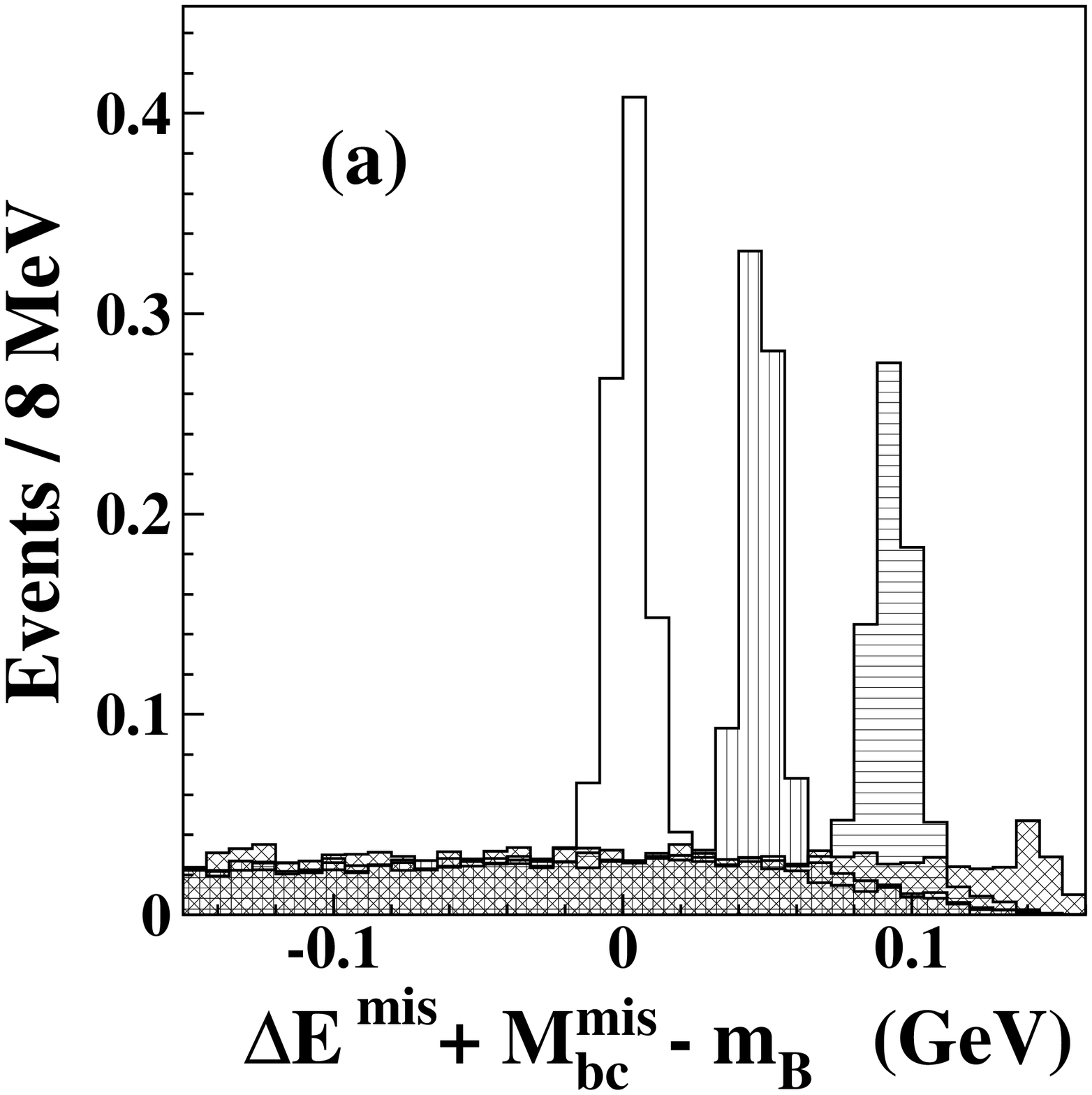,width=4.4cm,height=4.4cm}\epsfig{file=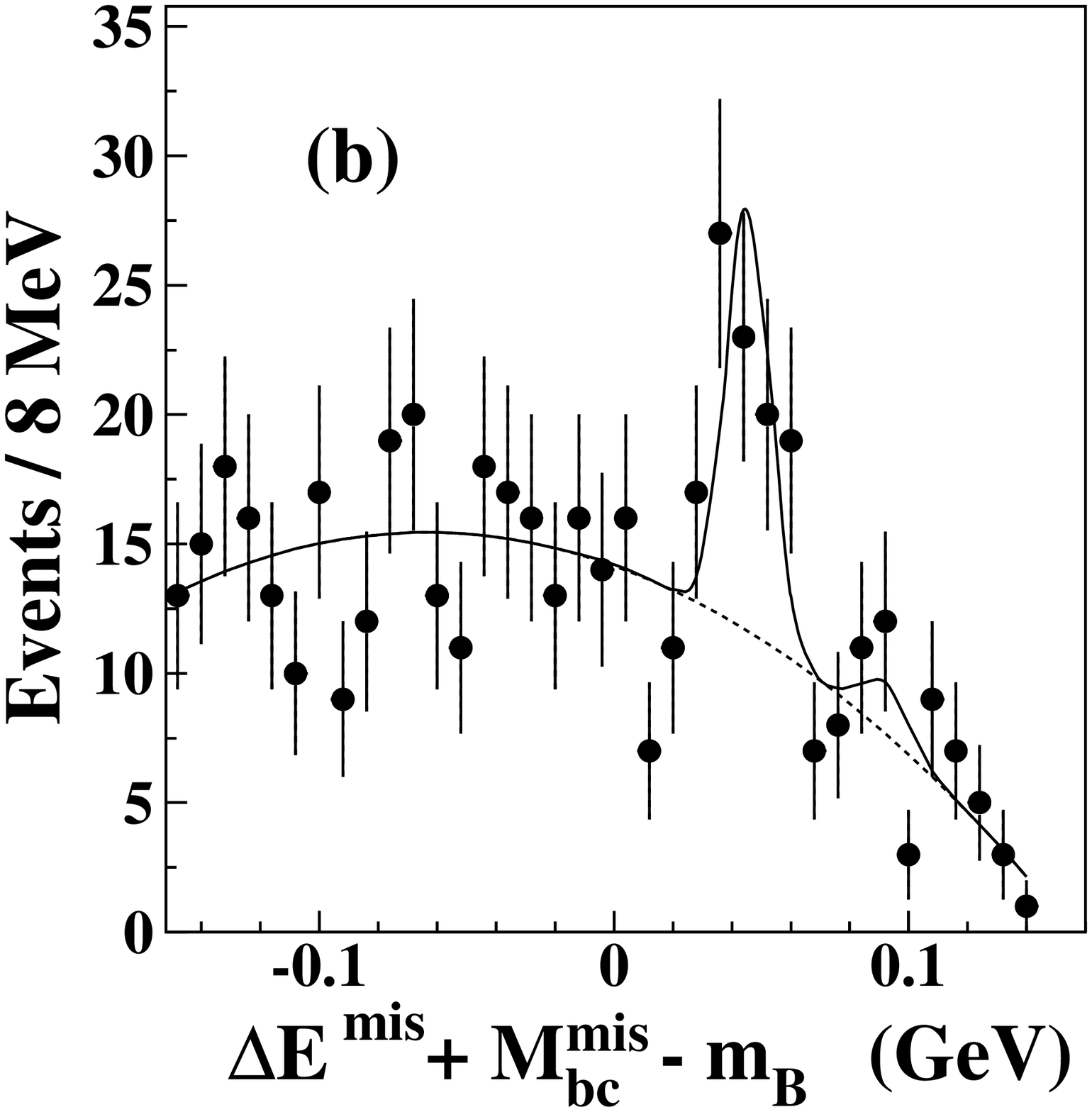,width=4.4cm,height=4.4cm}
\end{center}
\vspace{-0.6cm}
\caption{(a) The \mbox{$\Delta E^{\rm mis}+M_{\rm bc}^{\rm mis}-m_B$}
distribution normalized per reconstructed $B$ meson
for the MC simulated $B^+ \to J/\psi K^+$ decays
in the (peaks from left to right) $B\bar{B}\,\pi^+$,
$B\bar{B}^\ast\,\pi^+ +B^\ast\bar{B}\,\pi^+$,
$B^\ast\bar{B}^\ast\,\pi^+$, and $B\bar{B}\,\pi \pi$
channels.
(b) The \mbox{$\Delta E^{\rm mis}+M_{\rm bc}^{\rm mis}-m_B$} data
distribution for right-sign $B^{-/0}\,\pi^+$ combinations 
for all five studied $B$ modes. The curve shows the
result of the fit described in the text.}
\vspace{-0.2cm}
\end{figure}

The event yields and rates
are listed in Table III.
The three-body rates are calculated assuming the ratio of charged and neutral
directly produced pions to be 2:1 as expected from isospin conservation.
The pion reconstruction efficiencies are obtained from the three-body
phase-space matrix element MC simulation to be $73.8\%$,
$64.2\%$ and $57.2\%$ for 
the $B\bar{B}\,\pi^+$, $B\bar{B}^\ast\,\pi^+ +B^\ast\bar{B}\,\pi^+$,
and $B^\ast\bar{B}^\ast\,\pi^+$ channels, respectively.
In neutral $B$ modes, an efficiency correction is applied to take
into account the effect of 19$\%$ $B^0-\bar{B}^0$ mixing. 
The systematic uncertainties due
to the fit procedure ($\pm$1.5 signal events for each channel) 
and MC efficiency calculations (4--10$\,\%$) are also included
in the total systematic uncertainties.
The statistical significance of the 
$B\bar{B}^\ast\,\pi^+ +B^\ast\bar{B}\,\pi^+$ signal is 4.4$\sigma$.

It is interesting to note that 
by directly reconstructing pions, 
we observe only about one half of the rate obtained above for the
large $M_{\rm bc}$ region (Table II) resulting
in a deficit of $(9.2\,^{+3.0}_{-2.8} \pm 1.0)\,\%$.
If the difference were due to $B\bar{B}\,\pi\pi$ events,
we would expect $4 \pm 2$ events in the three rightmost bins
of the \mbox{$\Delta E^{\rm mis}+M_{\rm bc}^{\rm mis}-m_B$} distribution
of Fig.~3(b), where only one is observed.
In addition, the four-body channel is theoretically expected to have a rate
at least an order of magnitude smaller than the three-body channel rates,
because there is limited phase-space for the creation
of an additional pion.
Instead, we find that the deficit can be explained by ISR contributions
such as $e^+ e^- \to \Upsilon$(4S)$\gamma \to B\bar{B}\gamma$.
We calculate a probability of $\sim\,$10$\,\%$ for hard photon
emission \cite{isr} by the electron or positron beam with subsequent $B$
production, and estimate that $\sim$40$\,\%$ of such events are due to 
radiative return to the $\Upsilon$(4S) resonance. This estimate agrees with
the observed residual and explains the peaking structure on the right side
of Fig.~2(b), which is dominated by radiative return to
the $\Upsilon$(4S) or slightly higher energies.

\vspace{-0.3cm}
\renewcommand{\arraystretch}{1.3}
\begin{table}[h!]
\caption{The three-body channel yields and fractions.
The yields are obtained from
a fit to the \mbox{$\Delta E^{\rm mis}+M_{\rm bc}^{\rm mis}-m_B$} 
distribution using five studied $B$ decay modes.
The sum of the large $M_{\rm bc}$ channel rates is taken from
Table II. The residual is the difference between the sum
of the three-body channels and the result from the large $M_{\rm bc}$ region.}
\begin{center}
\begin{tabular}
{@{\hspace{0.05cm}}l@{\hspace{0.05cm}} @{\hspace{0.1cm}}c@{\hspace{0.1cm}} @{\hspace{0.1cm}}c@{\hspace{0.1cm}} @{\hspace{0.1cm}}c@{\hspace{0.05cm}}}
\hline \hline
 Channel & Yield ($\pi^+$), & Fraction over & Fraction per \\
  & events & large $M_{\rm bc}$ \ $\%$ & $b\bar{b}$ event \ $\%$ \\
\hline
$B\bar{B}\,\pi$ & $0.2\,^{+7.2}_{-6.9}$ & $0.2\,^{+6.8}_{-6.5}$ & $0.0 \pm 1.2 \pm 0.3$ \\
$B\bar{B}^\ast\pi+B^\ast\bar{B}\pi$ & $38.3\,^{+10.5}_{-\ 9.8}$ & $41.6\,^{+12.1}_{-11.4}$ & $7.3\,^{+2.3}_{-2.1} \pm 0.8$ \\
$B^\ast\bar{B}^\ast\,\pi$ & $4.8\,^{+6.4}_{-5.9}$ & $5.9\,^{+7.8}_{-7.2}$ & $1.0\,^{+1.4}_{-1.3} \pm 0.4$ \\
Residual &  & $52.3\,^{+15.9}_{-15.0}$ & $9.2\,^{+3.0}_{-2.8} \pm 1.0$ \\
Large $M_{\rm bc}$ & & 100. & $17.5\,^{+1.8}_{-1.6} \pm 1.3$ \\
\hline \hline
\end{tabular}
\end{center}
\vspace{-0.3cm}
\end{table}

We analyze other potential sources and backgrounds for the events in the 
multibody region.
The rate for a $b\bar{b}$ event to produce the $\Upsilon$(4S)
and two pions is expected to be less than 1$\,\%$ \cite{chen}.
The wide $J^P=1^+$ $B^{**}$ meson can be produced at the
$\Upsilon$(5S) CM energy with a subsequent decay resulting 
in three- or four-body 
channels, however this process is expected to be negligible
due to the very small phase-space.
The decay $B_s^0 \to B^+ e^- \bar{\nu_e}$ could 
contribute as a background to the multibody channels,
however, the corresponding branching 
fraction is estimated to be less than 10$^{-4}$.

In conclusion, the production of $B^+$ and $B^0$ mesons is measured
at the energy of the $\Upsilon$(5S).
Using fully reconstructed $B^+$ and $B^0$ mesons
the production rates per $b\bar{b}$ event are measured to be
$f(B^+) = (72.1 ^{+3.9}_{-3.8} \pm 5.0)\%$ and
$f(B^0) = (77.0^{+5.8}_{-5.6} \pm 6.1)\%$.
The average value $(73.7 \pm 3.2 \pm 5.1)\%$
agrees within uncertainties with
the CLEO value of $(58.9 \pm 10.0 \pm 9.2)\%$ \cite{cleob}.
Taking into account the $B_s^{(*)} \bar{B}_s^{(*)}$ event rate
at the $\Upsilon$(5S) of
$f_s = (19.5^{+3.0}_{-2.2})\%$ \cite{pdg}
(this value was obtained neglecting bottomonium,
resulting in an additional absolute uncertainty of about $1\%$),
and assuming the fraction of final states 
with a bottomonium meson to be
\mbox{$1 - f(B^+)/2 - f(B^0)/2 - f_s$},
some room for unobserved transitions still remains.

Assuming equal production of $B^+$ and $B^0$ mesons
we also measure 
the fractions for $b\bar{b}$ event transitions to
the two-body channels with $B^{+/0}$ meson pairs,
$f(B\bar{B}) = (5.5\,^{+1.0}_{-0.9} \pm 0.4)\,\%$,
$f(B\bar{B}^*+B^*\bar{B}) = (13.7 \pm 1.3 \pm 1.1)\,\%$,
$f(B^*\bar{B}^*) = (37.5\,^{+2.1}_{-1.9} \pm 3.0)\,\%$.
The $B\bar{B}$ channel is measured for the first time.
These fractions are in rough
agreement with theoretical predictions \cite{teoc,teod},
however further adjustment of the theoretical models
is required.

Using the additional charged pion directly produced in
$B^{(*)}\bar{B}^{(*)}\,\pi^+$ channels, we measure
the three-body channel fractions in a model-dependent way.
The $B\bar{B}^\ast\,\pi+B^\ast\bar{B}\,\pi$ decay channel
is observed for the first time with the fraction
$f(B\bar{B}^\ast\pi+B^\ast\bar{B}\pi) = (7.3\,^{+2.3}_{-2.1} \pm 0.8)\,\%$.
This measured three-body fraction is significantly larger
than those predicted in \cite{sim,lel}.
 
We thank the KEKB group for the excellent operation of the
accelerator, the KEK cryogenics group for the efficient
operation of the solenoid, and the KEK computer group and
the National Institute of Informatics for valuable computing
and SINET3 network support.  We acknowledge support from
the Ministry of Education, Culture, Sports, Science, and
Technology (MEXT) of Japan, the Japan Society for the 
Promotion of Science (JSPS), and the Tau-Lepton Physics 
Research Center of Nagoya University; 
the Australian Research Council and the Australian 
Department of Industry, Innovation, Science and Research;
the National Natural Science Foundation of China under
contract No.~10575109, 10775142, 10875115 and 10825524; 
the Ministry of Education, Youth and Sports of the Czech 
Republic under contract No.~LA10033 and MSM0021620859;
the Department of Science and Technology of India; 
the BK21 and WCU program of the Ministry Education Science and
Technology, National Research Foundation of Korea,
and NSDC of the Korea Institute of Science and Technology Information;
the Polish Ministry of Science and Higher Education;
the Ministry of Education and Science of the Russian
Federation and the Russian Federal Agency for Atomic Energy;
the Slovenian Research Agency;  the Swiss
National Science Foundation; the National Science Council
and the Ministry of Education of Taiwan; and the U.S.\
Department of Energy.
This work is supported by a Grant-in-Aid from MEXT for 
Science Research in a Priority Area (``New Development of 
Flavor Physics''), and from JSPS for Creative Scientific 
Research (``Evolution of Tau-lepton Physics'').

\end{document}